\documentclass[10pt,conference]{IEEEtran}
\IEEEoverridecommandlockouts
\usepackage{balance}
\usepackage{amssymb}
\usepackage{stmaryrd}
\usepackage{cite}
\usepackage{color}
\usepackage{multirow}
\usepackage{graphicx,times}
\usepackage{epstopdf}
\usepackage{indentfirst}
\usepackage{CJK}
\usepackage{amsmath}
\usepackage{amsfonts}
\usepackage{txfonts}
\usepackage{mathrsfs}
\usepackage{theorem}

\newtheorem{proposition}{Proposition}

\begin{document}
\title{On Secure Transmission Design: An Information Leakage Perspective}

\author{\IEEEauthorblockN{Yong Huang{$ ^\dagger $}, Wei Wang{${^\ast}{^\dagger}$}, Biao He{${^\ddagger}$}, Liang Sun{${^\S}$}, Tao Jiang{${^\dagger}$}}\IEEEauthorblockA{{${^\dagger}$}School of Electronic Information and Communications, Huazhong University of Science and Technology\\
		{${^\ddagger}$} Department of Electrical Engineering and Computer Science, University of California, Irvine\\{${^\S}$} School of Electronic and Information Engineering, Beihang University\\Email: \{yonghuang, weiwangw, taojiang\}@hust.edu.cn, biao.he@uci.edu, eelsun@buaa.edu.cn}
	\thanks{${^\ast}$The Corresponding author is Wei Wang (weiwangw@hust.edu.cn).}}

\maketitle

\begin{abstract}
Information leakage rate is an intuitive metric that reflects the level of security in a wireless communication system, however, there are few studies taking it into consideration. Existing work on information leakage rate has two major limitations due to the complicated expression for the leakage rate: 1) the analytical and numerical results give few insights into the trade-off between system throughput and information leakage rate; 2) and the corresponding optimal designs of transmission rates are not analytically tractable. To overcome such limitations and obtain an in-depth understanding of information leakage rate in secure wireless communications, we propose an approximation for the average information leakage rate in the fixed-rate transmission scheme. Different from the complicated expression for information leakage rate in the literature, our proposed approximation has a low-complexity expression, and hence, it is easy for further analysis. Based on our approximation, the corresponding approximate optimal transmission rates are obtained for two transmission schemes with different design objectives. Through analytical and numerical results, we find that for the system maximizing throughput subject to information leakage rate constraint, the throughput is an upward convex non-decreasing function of the security constraint and much too loose security constraint does not contribute to higher throughput; while for the system minimizing information leakage rate subject to throughput constraint, the average information leakage rate is a lower convex increasing function of the throughput constraint.
\end{abstract}

\maketitle
\vspace{-0.5cm}
\section{Introduction}
The inherent openness of wireless channel makes data transmission difficult to shield from unintended recipients \cite{wang2017securing}. Although the security of wireless transmission is safeguarded by traditional cryptographic techniques, an eavesdropper with a strong computational ability still can decipher confidential information from the received symbols by using brute-force attacks \cite{wang2018resonance}. In this situation, the transmitted information is completely leaked to the eavesdropper over wireless channel, which is the worst throughput-information leakage rate relation for securing wireless transmission. Physical layer security has been proposed in \cite{shannon1949communication} for ensuring secure wireless communications by exploiting the characteristics of wireless channels without any assumption on the computation capability of the eavesdropper. With the help of physical layer security, more confidential information could be transmitted on secure wireless transmission by introducing random noises into the transmitted messages to confuse the eavesdropper. Hence, it is of great interest to investigate the throughput-information leakage rate trade-off on secure transmission system. 

Although there are many secure transmission designs taking throughput into account, limited studies on information leakage rate are found in the literature. Specifically, \cite{he2016secrecy} proposed the average information leakage rate over quasi-static fading channels, which tells how fast on average the information is leaked to the eavesdropper, and examines average information leakage rate in secure transmission design. The major limitations in \cite{he2016secrecy} are: 1) the analytical and numerical results give few insights into the trade-off between system throughput and information leakage rate; 2) and the corresponding optimal designs of transmission rates are not analytically tractable. Besides, implicit exponential integral function is involved in the average information leakage rate, which hinders the further research on information leakage rate in wireless communication systems. 
 
In this work, we study the secure transmission design from an information leakage perspective. To address the above limitations, we reasonably approximate average information leakage rate and, based on the approximation, two transmission systems are considered. The major contributions of this paper are:
\begin{itemize}
	\item We thoroughly study the secure transmission design from an information leakage perspective. We propose a reasonable approximation for average information leakage rate. Based on the approximation, for the system maximizing throughput subject to information leakage rate constraint, we obtain the closed-form approximate optimal secure transmission rates; for the system minimizing information leakage rate subject to throughput constraint, we get the low-complexity approximate optimal secure transmission rates.
\end{itemize}
\begin{itemize}
	\item Based on the derived approximations, the throughput-information leakage rate trade-offs over wireless communication systems are comprehensively investigated. It is found that for the system maximizing throughput subject to information leakage rate constraint, the throughput is an upward convex non-decreasing function of security constraint. However, for the system minimizing information leakage rate subject to throughput constraint, the average information leakage rate is a lower convex increasing function of the throughput constraint. In the both systems, too stringent or too loose throughput constraints will suffer the systems.
\end{itemize}

The remainder of this paper is organized as follows. In section \uppercase\expandafter{\romannumeral 2}, we illustrate the secure transmission and problem formulation, and in section \uppercase\expandafter{\romannumeral 3} we detail the system design with the proposed approximation. Next, section \uppercase\expandafter{\romannumeral 4} shows the numerical results. Finally, the paper is concluded in section \uppercase\expandafter{\romannumeral 5}. 
\section{Secure Transmission and Problem Formulation}

We consider the wiretap-channel system, where a transmitter, Alice, sends confidential information to an intended receiver, Bob, in the presence of an eavesdropper, Eve. Alice, Bob and Eve are assumed to all have a single antenna. We refer to the Alice-Bob channel as the main channel and the Alice-Eve channel as the eavesdropper's channel. Both channels are assumed to undergo independent quasi-static fading. Then, the instantaneous channel capacity of Bob or Eve is given by
\begin{align}
C_i = \log_2 \left( 1 + \gamma_i \right),  i=b \; \text{or} \; e,
\end{align}
where $ \gamma_i $ denotes the instantaneous signal-to-noise ratios (SNRs), the subscripts $ b $ and $ e $ denote Bob and Eve, respectively. We adopt the quasi-static Rayleigh fading channel model, and the instantaneous SNR has an exponential distribution, which is given by
\begin{align}\label{subsec exponetial distribution}
f(\gamma_i )= \frac{1}{\bar{\gamma_i}} \exp\left(- \frac{\gamma_i}{\bar{\gamma_i}} \right), i=b \; \text{or} \; e,
\end{align}
where $ \bar{\gamma_i} $ denotes the average received SNR at Bob or Eve.

\subsection{Secure Encoding}
We consider the widely-adopted wiretap code \cite{wyner1975wire} for confidential message transmissions. Specifically, there are two transmission rates,
namely, the codeword transmission rate, $ R_b $, and the confidential information rate, $ R_s $. The rate cost for providing secrecy is defined as the positive rate difference between $ R_b $ and $ R_s $, which is expressed as  
\begin{align}
\phi = R_b - R_s.
\end{align}
The rate cost indicates the extra bits to introduce randomness for providing secrecy against the eavesdropper  \cite{jameel2017secrecy} in a codeword.
 A length $ n $ wiretap code is constructed by generating $ 2^{nR_{b}} $ codewords $ x^{n} (w,v) $, where $ w = 1,2,\cdots , 2^{n R_s} $ and $ v = 1,2,\cdots, 2^{ n (R_{b} - R_{s} ) } $. For each message index $ w $, we randomly select $ v $ from $ \left\lbrace 1,2,\cdots ,2^{ n (R_{b} - R_{s} ) }  \right\rbrace  $ with uniform probability and transmit the codeword $ x^{n} (w,v) $.

\subsection{Transmission Scheme}
 In this paper, we consider a fixed-rate transmission scheme, where the transmission rates, i.e., $ R_b $ and $ R_s $, are fixed over time. Bob and Eve are assumed to perfectly know their own channels. Hence, the values of $ C_b $ and $ C_e $ are known by Bob and Eve, respectively. Alice has the statistical knowledge of Bob and Eve's channels but does not know about the instantaneous channel capability. We further assume that Bob provides a one-bit feedback about his channel quality to Alice in order to avoid unnecessary transmissions  \cite{zhou2011rethinking,he2013secure}. The one-bit feedback enables an on-off transmission scheme to avoid connection outage. It is assumed that the transmission takes place only when $ R_b \leq C_b $ \cite{he2016secrecy}. Therefore, the on-off transmission scheme incurs a probability of transmission, which is given by
\begin{align}
p_{tx} = \mathbb{P} \left(  R_b \leq C_b  \right)  = \exp \left( - \frac{2^{R_b}-1}{\bar{\gamma_b}} \right).
\end{align}

\subsection{Metrics}
We adopt the average information leakage rate and the throughput to measure the secrecy performance and the rate performance of the system, respectively. 

Average information leakage rate is derived from fractional equivocation $ \Delta $ and confidential information rate $ R_s $, which is given by \cite{he2016secrecy}
\begin{align} \label{sec2:average_information_leakage_rate}
R_L & = \mathbb{E} \left\lbrace \left( 1- \Delta \right)R_s  \right\rbrace.
\end{align}
Therein, the fractional equivocation $ \Delta $ characterizes the level at which the eavesdropper is confused \cite{leung1978gaussian} because of the dynamism of the wireless channel. Thus, the average information leakage rate tells how fast the information is leaked to the eavesdropper. In the fixed-rate on-off transmission system, \eqref{sec2:average_information_leakage_rate} can be further derived as \cite{he2016secrecy}
\begin{align} 
R_L  = \frac{1}{\ln2} \exp\left( \frac{1}{\bar{\gamma_e}} \right) \left( Ei\left( - \frac{2^{R_b}}{\bar{\gamma_e}} \right)  - Ei\left( - \frac{2^{R_b - R_s}}{\bar{\gamma_e}} \right) \right),
\end{align}
where $ Ei(x) = \int_{-\infty}^{x} \frac{\exp\left( t\right) }{t} dt $ is the exponential integral function. We can find that $ R_L $ is increasing on $ R_s $ and decreasing on $ R_b $.

Throughput is a widely-adopted metric to measure the confidential information received by an intended user in a real-world communication system. In our system,  the throughput is given by
\begin{align}\label{sec3:throughput}
\eta = p_{tx} R_s = \exp\left(  - \frac{2^{R_b}-1}{\bar{\gamma_b}}\right)  R_s .
\end{align}  
Note that $ \eta $ is increasing on $ R_s $ and decreasing on $ R_b $.

\subsection{Cases}
To thoroughly understand the trade-off between throughput and information leakage rate, we consider the following two cases in secure transmission designs:
\begin{itemize}
	\item \textbf{Case 1}: Maximizing throughput with average information leakage rate constraint is considered. This case represents the scenario where the throughput of the transmission is given the priority and the security is secondary to be considered. 
\end{itemize}
\begin{itemize}
	\item \textbf{Case 2}: Minimizing average information leakage rate with throughput constraint is considered. This case represents the scenario where the security of the wireless link is first taken into consideration and the throughput is subordinate.
\end{itemize}
The above two cases summarize many widely-used wireless communication services like video call and mobile payment in practice, thus are of great significance to consider.    

\section{System Design with Approximation}
The implicit exponential integral function that is involved in average information leakage rate hinders the analytic analysis and incurs a high computation complexity in obtaining the optimal transmission rates of secure transmission. Although there are many existing works on approximations for the exponential integral function, none of them works in secure transmission designs directly. Consequently, an easy-to-evaluate approximation is desired for further research on the throughput-information leakage rate trade-off. To tackle this obstacle and examine the role of information leakage rate in wireless communication systems, in this section, we firstly propose a reasonable approximation for average information leakage rate. Subsequently, we detail optimal secure transmission design in the above two cases with the obtained approximation. At last, based on the approximation, analytical analysis is given at the end of each case.

\subsection{Approximation for the Average Information Leakage Rate}
To settle the obstacle brought by exponential integral function, we propose an approximation for average information leakage rate with the help of Abramowitz and Stegun's upper and lower bounds of $ Ei(\cdot) $ \cite{abramowitz1964handbook} and classical Lagrange's mean value theorem  \cite{rudin1976principles}. We denote $ R_{Lp} $ as the approximation for $ R_L $ when $ \bar{\gamma_b} $ is large. Note that large $ R_b $ will contribute to a high transmission rate of confidential information or high secrecy cost for preventing confidential information leaked. Thus, consideration of large $ R_b $ is of practical interest for the study on physical layer security. The approximation is summarized into the following proposition.
\begin{proposition}
	Approximation for the average information leakage rate, when $ \bar{\gamma_b} $ is large, is given by
	\begin{align} \label{sec3:approximation}
	R_{Lp} =  \frac{3 \bar{\gamma_e} \exp\left( \frac{1}{\bar{\gamma_e}}\right) }{10 \ln2} \frac{2^{R_s}-1}{2^{R_b}}.
	\end{align}
\end{proposition}
\begin{IEEEproof}
	See Appendix~A.
\end{IEEEproof}

As shown in \eqref{sec3:approximation}, $ R_{Lp} $ is the simple combination of basic functions and does not involve the exponential integral function $ Ei(\cdot) $, which makes it much easier to use and analyze. Based on the derived approximation, we can now derive the closed-form approximate optimal secure transmission rates in the following two different cases.

\subsection{Approximate the Optimal Transmission Rates for Case 1}\label{sec3:case1}
In Case 1, the system maximizing throughput subject to average information leakage rate constraint is considered, and the optimization problem is formulated as
\begin{align}\label{sec3:case1_orig_pro}
\mathop{\max } \limits_{R_b , R_s} & \quad   \eta = \exp\left(  - \frac{2^{R_b}-1}{\bar{\gamma_b}}\right)  R_s \: , \\
s.t. & \quad R_L\leq \xi \:, \: 0 \leq R_s \leq R_b,
\end{align}
where $ \xi  $ is the maximum average information leakage rate permitted in this transmission system. 

\underline{Problem reformulation.} 
With the approximation in Proposition 1, an approximate optimization problem for Case 1 is obtained, which is given by
\begin{align}\label{opt_problem}
\mathop{\max } \limits_{R_b , R_s} & \quad  \eta  = \exp\left(  - \frac{2^{R_b}-1}{\bar{\gamma_b}}\right)  R_s \: , \\
 s.t. & \quad R_{Lp}  \leq \xi \: , \: 0 \leq R_s \leq R_b \:.
\end{align}

\underline{Feasibility of constraint $ \xi $.} To obtain the feasible range of $ \xi $ is equivalent to determine the maximum average information leakage rate. Given any $ R_b $, we find that $ \partial R_{Lp} / \partial R_s $ is more than 0. Hence given any $ R_b $, it is wise to have maximum $ R_s $, i.e., $ R_s=R_b $, for maximizing $ R_{Lp} $. Then by having $ R_b $ approaching $ +\infty $, the upper bound of $ R_{Lp} $ is obtained. Thus, feasible range of $ \xi $ is given by
\begin{align}
0< \xi \leq  \frac{3 \bar{\gamma_e} \exp\left( \frac{1}{\bar{\gamma_e}} \right) }{10 \ln 2} .
\end{align} 

\underline{Approximate optimal transmission rates.} We denote $ W_0 (\cdot) $ as the principal branch of the Lambert W function and $ A= \frac{10 \ln2 \exp\left( - \frac{1}{\bar{\gamma_e}}\right) }{3 \bar{\gamma_e}} $. The closed-form optimal secure transmission rates for the approximate optimization problem are summarized into the following proposition.
\begin{proposition}\label{sec3:case1_proposition}
	Approximate optimal transmission rates of the system maximizing throughput subject to average information leakage rate constraint are given by
    \begin{align}\label{sec3:case1_opt_rs}
	R^{*}_{s1}=\log_2\left( 1+ \xi A 2^{R^{*}_{b1}} \right) 
	\end{align}
	and 
	\begin{align}\label{sec3:case1_opt_rb}
	R^{*}_{b1}= \max \left( R_{b1,min} \:,\: R_{b1,0} \right) ,
	\end{align}
	where 
	\begin{align}\label{sec3:case1_lower_bound_rb}
	R_{b1,min}=-\log_2\left( 1- \xi A \right)
	\end{align}
	and
	\begin{align}\label{sec3:case1_rb0}
	R_{b1,0}=\log_2 \left( \frac{\exp \left( W_{0} \left( \xi A \bar{\gamma_b} \right)  \right) -1}{\xi A} \right).
	\end{align} 
\end{proposition} 
\begin{IEEEproof}
	See Appendix~B.
\end{IEEEproof}

\underline{Analysis based on Proposition 2.} From Proposition 2, important observations could be obtained. When $ \xi = 0 $, we find that $ R^{*}_{s1}= 0 $, which implies no confidential information is transmitted if no information leakage is permitted. In this situation, the communication system is meaningless because only random noises is transmitted. Because $ R_{b1,min} $ is increasing and $ R_{b1,0} $ is decreasing with respect to $ \xi $, respectively, $ R_{b1,min} $ and $ R_{b1,0} $ may encounter at a certain value of $ \xi $, $ \xi_0 $, where $ \xi_0 $ is the solution of the equation $ R_{b1,0}= R_{b1,min} $. When $ \xi $ exceeds $ \xi_0 $, we have $ R^{*}_{b1}= R_{b1,min} $ and $ R^{*}_{s1} = R^{*}_{b1} $, which leads to zero secrecy cost. In this situation, the secure system is incapable, for no confidential information is shield from the eavesdropper over the wireless channel. The observations reveals that both too stringent and too loose security constraints will result in impractical communication systems.

\subsection{Approximate the Optimal Transmission Rates for Case 2}
In Case 2, the system minimizing average information leakage rate subject to throughput constraint is taken into account. The optimization problem is formulated as
\begin{align}\label{sec3:case2_orig_pro}
\mathop{\min }\limits_{R_s, R_b} & \; R_L =  \frac{1}{\ln2} \exp\left( \frac{1}{\bar{\gamma_e}} \right) \left( Ei\left( - \frac{2^{R_b}}{\bar{\gamma_e}} \right)  - Ei\left( - \frac{2^{R_b - R_s}}{\bar{\gamma_e}} \right) \right) \: ,\\
s.t. & \quad \eta \geq \Gamma \:, \: 0 \leq R_s \leq R_b \:,
\end{align}
where $ \Gamma $ represents the minimum throughput required in this transmission system.

\underline{Problem reformulation.} 
With the approximation in Proposition 1, an approximate optimization problem for Case 2 is obtained, which is given by
\begin{align}\label{sec3:case2_approxiamte_opt_pro}
\mathop{\min } \limits_{R_s, R_b} & \quad R_{Lp}  =  \frac{3 \bar{\gamma_e} \exp\left( \frac{1}{\bar{\gamma_e}}\right)  }{10 \ln2}   \frac{2^{R_s}-1}{2^{R_b}} \: ,\\
 s.t. & \quad \eta \geq \Gamma \: , \: 0 \leq R_s \leq R_b \:.
\end{align} 

\underline{Feasibility of constraint $ \Gamma $.} For fixed $ R_b $, we find that $ \partial \eta / \partial R_s $ is always positive. Hence for fixed $ R_b $, we have the maximum $ R_s $, i.e., $ R_s= R_b $, for maximizing $ \eta $. Then, by solving the equation $ \frac{\partial \eta(R_s =R_b)}{\partial R_b} =0$, the $ R_b $ maximizing $ \eta\left( R_s =R_b \right)  $ is obtained, which is equal to $ \frac{W_o(\bar{\gamma_b})}{\ln2} $. Thus, the feasible range of $ \Gamma $ is given by
\begin{align}
0\leq \Gamma \leq \frac{W_o(\bar{\gamma_b})}{\ln2} \exp\left( \frac{1-2^{\frac{W_o(\bar{\gamma_b})}{\ln2}}}{\bar{\gamma_b}}\right) .
\end{align}

\underline{Approximate optimal transmission rates.} We denote $ R_{b2,min} $ and $ R_{b2,max} $ as the solutions of $ x  $ to $ \exp\left( \frac{1-2^x}{\bar{\gamma_b}}\right)  x = \Gamma $, where $ R_{b2,min} < R_{b2,max} $. The optimal transmission rates for the approximate optimization problem are summarized into the following proposition. 
\begin{proposition}
Approximate optimal transmission rates of the system minimizing average information leakage rate subject to throughput constraint are given by
\begin{align}\label{sec3:case2_opt_rs}
R^{*}_{s2}=\Gamma \exp\left( \frac{2^{R^{*}_{b2}}-1}{\bar{\gamma_b}}\right) 
\end{align}
and 
\begin{align}\label{sec3:case2_opt_rb}
 R^{*}_{b2}=\left\lbrace 
\begin{array}{lcl}
{R_{b2,min}} \: , & \text{if} & R_{b2,0} < R_{b2,min} \: , \\
{R_{b2,0}} \: , & \text{if} & R_{b2,min} \leq R_{b2,0} \leq R_{b2,max} \: , \\
{R_{b2,max}} \: , & \text{if} & R_{b2, max} < R_{b2,0} \: ,
\end{array}
\right.
\end{align} 
where
\begin{align} \label{sec3:case2_r0}
R_{b2,0}=\log_2 \left( 1+\bar{\gamma_b} \ln\left( \frac{\ln B}{\Gamma \ln2} \right) \right) 
\end{align}
and $ B $ is the solution of $ x $ to the equation
\begin{align}\label{sec3:case2_r0_2}
x \ln\left( x\right)  \ln\left( \frac{1}{\Gamma \ln2 } \exp\left( \frac{1}{\bar{\gamma_b}} \right) \ln\left( x\right)   \right) -x +1 =0 \: .
\end{align}
\end{proposition}
\begin{IEEEproof}
	See Appendix~C.
\end{IEEEproof}

\underline{Analysis based on Proposition 3.} With the optimal secure transmission rates given in Proposition 3, insightful observations could be obtained. When $ \Gamma=0 $, we have $ R^{*}_{s2}=0 $, which implies that all the codeword transmission bits are leveraged to provide secrecy against the eavesdropper. In this situation, the system is unprofitable for no useful information is transmitted. When $ \Gamma =  \frac{W_o(\bar{\gamma_b})}{\ln2}$, we find that $ R^{*}_{s2} = R^{*}_{b2} $, which demonstrates that the secrecy cost is maximum for any given $ R_b $ and all the codeword transmission bits are occupied by the confidential information. In this situation, the system is unsafe because no confidential information is shield from eavesdropping over the wireless channel. The observations also tell that both too stringent and too loose throughput constraint will make the communication system impractical.

\section{Numerical Results}
 
In this section, we present the numerical results for wireless systems with $ \bar{\gamma_e}=3 \text{dB} $ and different levels of $ \bar{\gamma_b} $ to demonstrate the role of information leakage on secure transmission designs. In each figure, the exact curve is obtained by numerically solving the optimization problem \eqref{sec3:case1_orig_pro} or \eqref{sec3:case2_orig_pro}, however, the approximate curve is obtained by Proposition 2 or 3.

\begin{figure}
	\centering
	\includegraphics[width=0.85\linewidth]{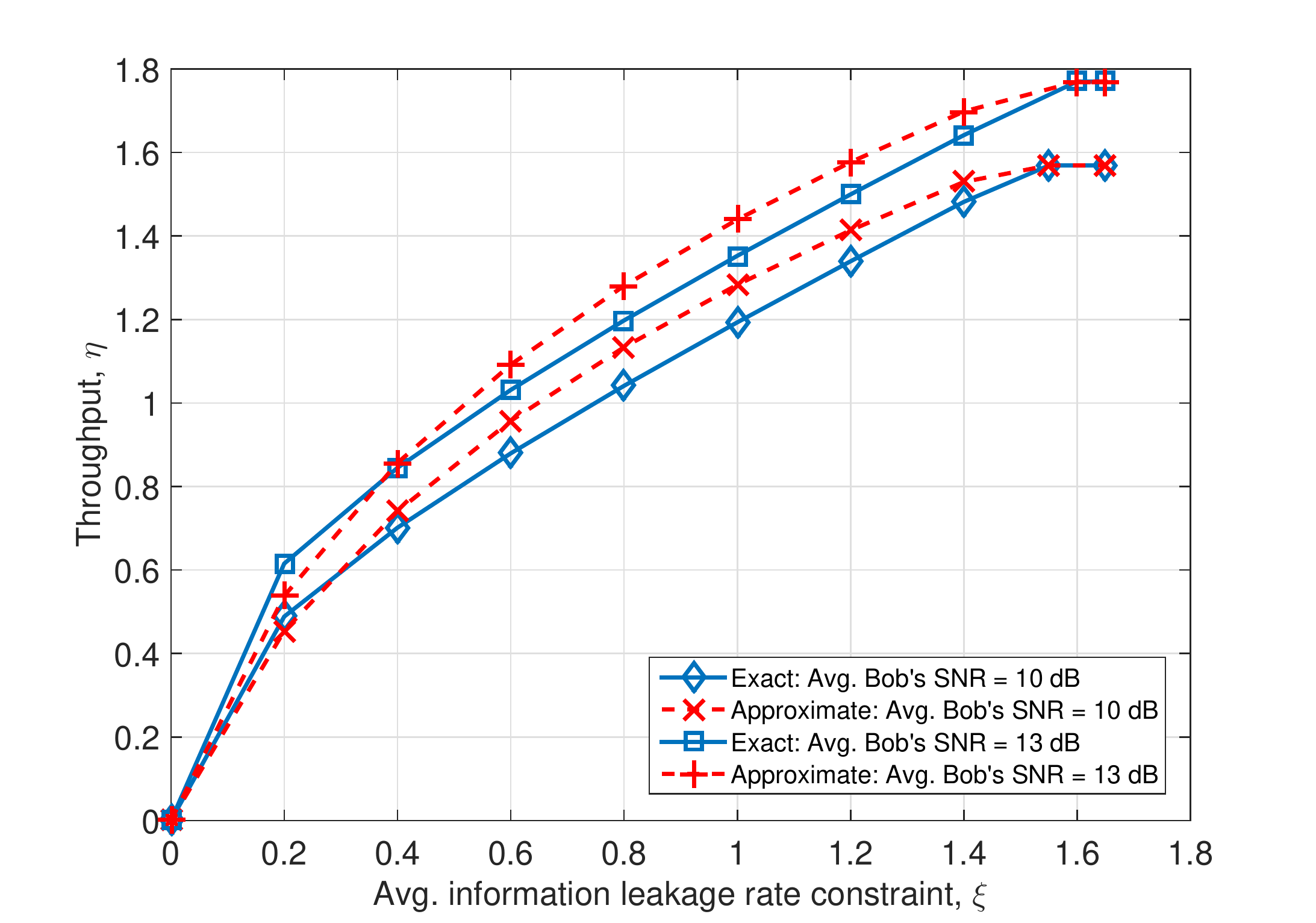}
	\caption{: Throughput versus average information leakage rate constraint with $ \bar{\gamma_e} = 3 \text{dB} $ in Case 1.}
	\label{fig:figure1}
\end{figure}

We first present throughput versus security constraint subject to different levels of Bob's average SNR in Case 1. Fig.~\ref{fig:figure1} shows that $ \eta $ is an upward convex non-decreasing function of $ \xi $. And there exists a limitation on the throughput as the security constraint increases, which means much too loose security constraint does not contribute to higher throughput in this case. Another observation is that the system with a higher $ \bar{\gamma_b} $ achieves higher throughput. But when the security constraint is very small, the difference in throughput between the systems is small. It is worth noting that the observations above are founded on both the exact and approximate curves. In addition, the maximum absolute error of the approximate curves with $ \bar{\gamma_b} = 10 \text{dB} $ and $ \bar{\gamma_b} = 13 \text{dB} $ are less than 0.2 and 0.1, respectively, which indicates the approximation of $ Ei(\cdot) $ is reasonable in Case 1.

\begin{figure}
	\centering
	\includegraphics[width=0.85\linewidth]{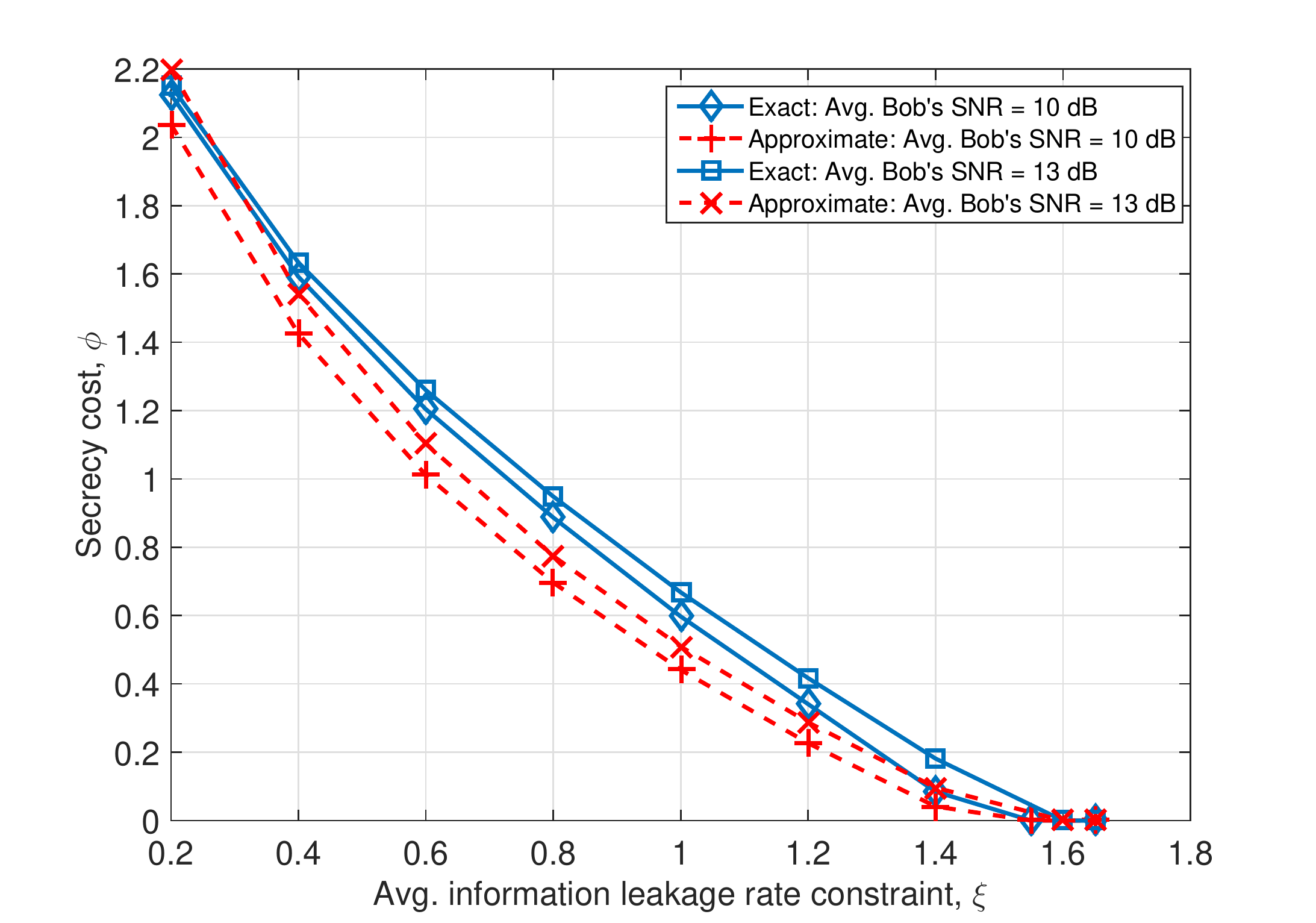}
	\caption{: Transmission rates versus average information leakage rate constraint with $ \bar{\gamma_b}=13 \text{dB} $ and $ \bar{\gamma_e} = 3 \text{dB} $ in Case 1.}
	\label{fig:figure2}
\end{figure}

We then focus on secrecy cost versus security constraint for different systems in Case1. As depicted in Fig.~\ref{fig:figure2}, $ \phi $ is a non-increasing function of $ \xi $. And there exists a certain value, beyond which the secrecy cost equals zero as the security constraint increases. This observation confirms our analytic analysis that when $ \xi $ exceeds $ \xi_0 $, $ R^{*}_{s1} = R^{*}_{b1} $. We also find that the system with better main channel quality pays a higher secrecy cost against eavesdropping. We can see these observations found on both the exact and approximate curves. Additionally, the maximum absolute distances between the exact and approximate curves with $ \bar{\gamma_b}=10 \text{dB} $ and $ \bar{\gamma_b}=13 \text{dB} $ are less than 0.25 and 0.2, respectively.

\begin{figure}
	\centering
	\includegraphics[width=0.85\linewidth]{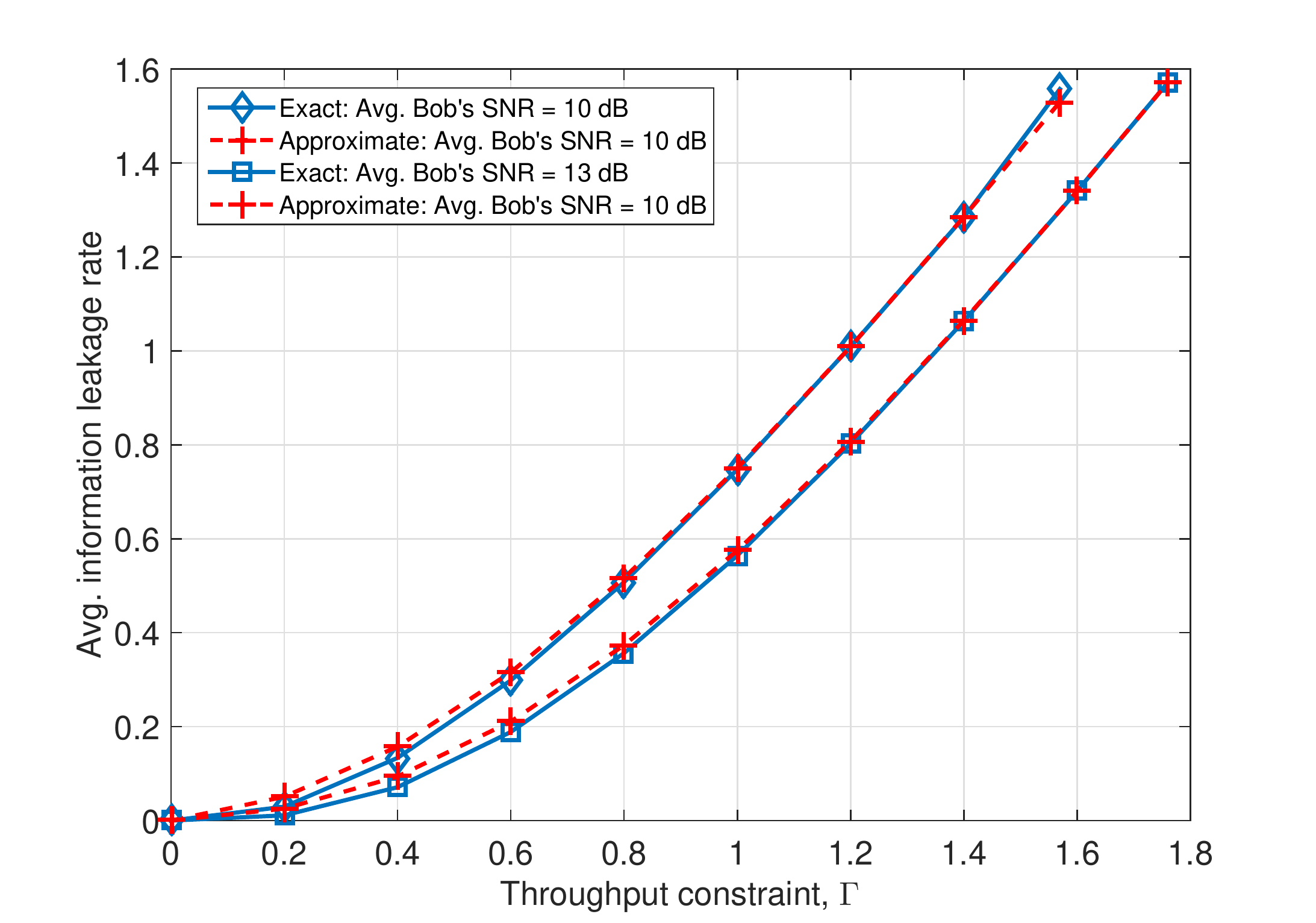}
	\caption{: Average information leakage rate versus throughput constraint with $ \bar{\gamma_e} = 3 \text{dB} $ in Case 2.}
	\label{fig:figure3}
\end{figure}

Next, we illustrate average information leakage rates versus throughput constraint subject to different levels of Bob's average SNR in Case 2. As shown in Fig.~\ref{fig:figure3}, average information leakage rate is a lower convex increasing function of $ \Gamma $. However, there is no limitation on information leakage rate like throughput in Case 1. We also observed that the system with a higher $ \bar{\gamma_b} $ has a lower average information leakage rate. However, when the throughput constraint is very small, the difference in average information leakage rate between the systems is small. Note that the maximum absolute error of $ R_{Lp} $ with $ \bar{\gamma_b} = 10 \text{dB} $ and $ \bar{\gamma_b} = 13 \text{dB} $ are less than 0.1 and 0.05, respectively, which indicates the approximation is reasonable in Case 2, too.

\begin{figure}
	\centering
	\includegraphics[width=0.85\linewidth]{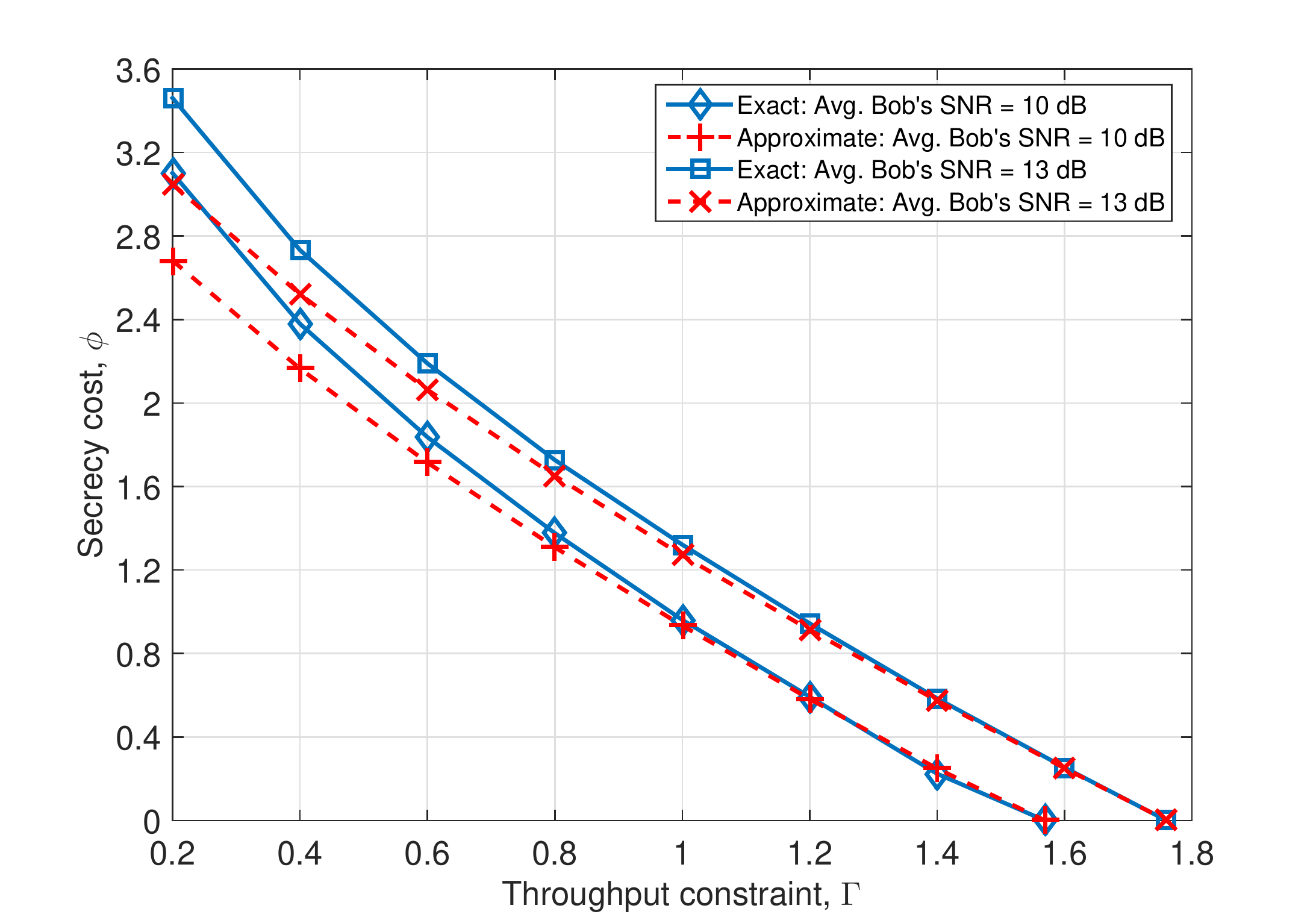}
	\caption{: Transmission rates versus throughput constraint with $ \bar{\gamma_b}=13\text{dB} $ and $ \bar{\gamma_e} = 3 \text{dB} $ in Case 2.}
	\label{fig:figure4}
\end{figure}

Finally, we show secrecy cost versus throughput constraint for different communication systems in Case 2. As depicted in Fig.~\ref{fig:figure4}, the secrecy cost decreases as the throughput constraint increases. And the secrecy cost goes to zero at the upper bound of the throughput constraint. The observation testifies our analytical analysis that when $ \Gamma =  \frac{W_o(\bar{\gamma_b})}{\ln2}$, $ R^{*}_{s2} = R^{*}_{b2} $. We also observe that the system with a high average main channel SNR obtains more secrecy cost for more random noises in the codeword transmission rate needed to disturbing the eavesdropper for secure transmission. Note that the maximum absolute distances between exact and approximate curves in two systems are both less than 0.4.

\section{Conclusion}
In this work, we investigate secure transmission designs from an information leakage perspective. Specifically, to address the obstacle brought by exponential integral function, we proposed a reasonable approximation for average information leakage rate. Based on the proposed approximation, the optimal transmission rates in two cases are obtained, respectively. Through analytical and numerical results, we found that for the system maximizing throughput subject to information leakage rate constraint, the throughput is an upward convex non-decreasing function of the security constraint. However, for the system minimizing information leakage rate subject to throughput constraint, the average information leakage rate is a lower convex increasing function of the throughput constraint. Additionally, in each case, a higher average Bob's SNR contributes to better throughput or security performance over the wireless channel. It is worth noting that all insightful observations are found on both the exact and approximate curves and the maximum absolute error in all figures are low, which testifies the proposed approximation is reasonable. Altogether, these results provide some important insights on information leakage in secure transmission designs.

\begin{appendices} 

\section{Proof of Proposition 1} 
From $ R_L =  \frac{1}{\ln2} \exp\left( \frac{1}{\bar{\gamma_e}} \right) \left( Ei\left( - \frac{2^{R_b}}{\bar{\gamma_e}} \right)  - Ei\left( - \frac{2^{R_b - R_s}}{\bar{\gamma_e}} \right) \right) $, we find that to approximate $ R_L $ is mainly to approximate
\begin{align}
O = Ei\left( - \frac{x}{\bar{\gamma_e}} \right)  - Ei\left( - \frac{x }{\bar{\gamma_e} y} \right),
\end{align} 
where $ x$ and $ y $ denote $ 2^{R_b} $ and $ 2^{R_s} $, for simplicity, respectively. With Abramowitz and Stegun's bounds  of $ Ei(\cdot) $ \cite{abramowitz1964handbook}, we have
\begin{align} \label{sec3:sub1_O}
O \approx  & -\frac{1}{2} \exp\left( -\frac{x}{\bar{\gamma_e}}\right)  \ln\left( 1+\frac{2\bar{\gamma_e}}{x} \right)+ \exp\left( -\frac{x}{\bar{\gamma_e} y}\right)  \ln\left( 1+\frac{\bar{\gamma_e}y}{x} \right) .
\end{align}

With infinitesimal equivalence, $ \ln\left( 1+ \frac{2\bar{\gamma_e}}{x} \right) \approx  \frac{2\bar{\gamma_e}}{x} $ and	$ \ln\left(1+ \frac{\bar{\gamma_e} y}{x} \right) \approx \frac{\bar{\gamma_e} y}{x} $ when $ \bar{\gamma_b} $ is large. By doing the corresponding equivalence infinitesimal replacement in \eqref{sec3:sub1_O}, we have
\begin{align}
\notag O \approx & -\frac{1}{2}\exp\left( -\frac{x}{\bar{\gamma_e}}\right)  \frac{2\bar{\gamma_e}}{x}+ \exp\left( -\frac{x}{\bar{\gamma_e} y}\right)  \frac{\bar{\gamma_e}y}{x}\\
\notag  = & \frac{\bar{\gamma_e}}{x} \left( y \exp\left( - \frac{x}{\bar{\gamma_e} y}\right)  - \exp\left( -\frac{x}{\bar{\gamma_e}}\right) \right)\\ 
= & \bar{\gamma_e} \frac{y-1}{x}  \left( \frac{y \exp\left( -\frac{x}{\bar{\gamma_e}y}\right) - \exp\left( -\frac{x}{\bar{\gamma_e}}\right) }{y-1} \right) .
\end{align}
We denote $ f(z)=z \exp\left( -\frac{x}{\bar{\gamma_e}z}\right)  $ and $ f(z) $ with respect to $ z $ is continuous and differentiable on closed interval $ \left[ 1 ,y \right]  $. From Lagrange's mean value theorem \cite{rudin1976principles}, there exists a point $ c $ in $ (1 \: , \: y) $ such that $ f^{'}(c)=\frac{f(y)-f(1)}{y-1} $. Hence, we have 
\begin{align}
O \approx f^{'}(c) \bar{\gamma_e} \frac{y-1}{x}.
\end{align}
Besides, when $ 1 \leq x \leq \infty $ and $ 1 \leq z \leq x $, we have $ 0 < f^{'}(z) <1 $. By setting $ \alpha = f^{'}(c) $, we obtain
\begin{align}
O  \approx \alpha \bar{\gamma_e} \frac{y-1}{x},
\end{align}
where $ \alpha \in \left(0 \: , \: 1 \right)  $. To reduce computational complexity in our problem, we empirically set $ \alpha=3/10 $, which provides good accuracy. Recalling that $ x=2^{R_b} $ and $ y=2^{R_s} $, we have the approximation in \eqref{sec3:approximation}. This completes the proof.

\section{Proof of Proposition 2} 
	From the constraint condition $ R_{Lp} \leq \xi $, we have
\begin{align}
R_s \leq \log_2\left( 1+ \xi A 2^{R_b} \right).
\end{align}
Given any $ R_b $, we find that $ \partial \eta / \partial R_s $ is always more than 0. Hence given any $ R_b $, it is wise to have the maximum $ R_s $, i.e., $ R_s = \log_2\left( 1+ \xi A 2^{R_b} \right) $, for maximizing $ \eta $. Thus, $ R^{*}_{s1} $ is obtained in \eqref{sec3:case1_opt_rs}. Additionally, with $ R_s = \log_2\left( 1+ \xi A 2^{R_b} \right) $ and $ R_s \leq R_b $, the feasible range of $ R_b $ is given by 
\begin{align}
R_{b1,min}=-\log_2\left( 1- \xi A \right) \leq R_b.
\end{align}
Then, we can rewrite the optimization problem as
\begin{align}
\mathop{\max } \limits_{R_b} & \quad  \eta|_{R_s=\log_2\left( 1+ \xi A 2^{R_b}  \right)} \: , \label{sec3:case1_rewirite_opt_pro}  \\
s.t & \quad  R_{b1,min} \leq  R_b  \: .\label{sec3:case1_rewirite_opt_pro_condition}
\end{align}
Finally, by solving for $ R_s $ in the equation $ \frac{\partial \eta(R_b)}{\partial R_b} =0 $, the only maximum value point $ R_{b1,0} $ can be obtained in \eqref{sec3:case1_rb0}. Hence, $ R^{*}_{b1} $ is obtained in \eqref{sec3:case1_opt_rb}. This completes the proof.

\section{Proof of Proposition 3} 
From the constraint condition $ \eta \geq \Gamma $, we have 
\begin{align}
R_s \geq \Gamma \exp\left( \frac{2^{R_b}-1}{\bar{\gamma_b}}\right) .
\end{align}
Given any $ R_b $, we find that $ \partial R_{Lp} / \partial R_s $ is always more than 0. Hence given any $ R_b $, it is wise to have the minimum $ R_s $, i.e., $ R_s=\Gamma \exp\left( \frac{2^{R_b}-1}{\bar{\gamma_b}}\right)  $, for minimizing $ R_{Lp} $. Thus, $ R^{*}_{s2} $ is given in \eqref{sec3:case2_opt_rs}. 

Then, as analyzed in obtaining the feasibility of $ \Gamma $, given any $ R_b $, it is wise to have the maximum $ R_s $, i.e., $ R_s  =  R_b $, for maximizing $ \eta $. Hence, we can obtain the feasible range of $ R_b $ for satisfying the throughput constraint by solving $ R_s $ in the equation $ \eta (R_s = R_b) = \Gamma $. The feasible range is given by
\begin{align}
R_{b2,min} \leq R_b \leq R_{b2,max}.
\end{align}

Consequently, the optimization problem can be rewrite as 
\begin{align}
\mathop{\min }\limits_{R_b} & \quad R_{Lp}|_{R_s=\Gamma \exp\left( \frac{2^{R_b}-1}{\bar{\gamma_b}}\right) } \: , \label{sec3_case2_rewrite_opt_pro} \\
s.t. & \quad R_{b2,min} \leq R_b \leq R_{b2,max} \: .
\end{align}

Finally, the minimum value point $ R_{b2,0} $ can be obtained by solving for $ R_b $ in the equation $\frac{\partial R_{Lp} ( R_b )}{\partial R_b} = 0 $. We find that the closed-form solution of $ R_{b2,0} $ is mathematically intractable. We can numerically obtain $ R_{b2,0} $ by \eqref{sec3:case2_r0} and \eqref{sec3:case2_r0_2}. Thus, $ R^{*}_{b2} $ is given in \eqref{sec3:case2_opt_rb}. This completes the proof.

\end{appendices} 

\section*{Acknowledgement}
The research was supported in part by the National Science Foundation of China under Grant 61502114 and 91738202, as well as the Fundamental Research Funds for the Central Universities, HUST: 2018JYCXJJ031.

\bibliographystyle{IEEEtran}
\bibliography{IEEEabrv,./Leakage_rate}

\end{document}